\documentclass[aps,prd,superscriptaddress,nofootinbib,preprint]{revtex4-1}
\usepackage[utf8]{inputenc}
\usepackage{ulem}
\usepackage{color}
\usepackage{graphicx}
\usepackage{multirow}
\usepackage{amsmath,amssymb,amsthm,amsxtra,overpic,bbm,bm,float
,epsfig}

\newcommand{\tempbf}{}
\usepackage{color}

\fontsize{3}{5}\selectfont

\begin{document}

\title{Extracting nuclear form factors with coherent neutrino scattering}
\author{Emilio Ciuffoli} 
\affiliation{Institute of Modern Physics, CAS. NanChangLu 509, Lanzhou 730000, China}
\author{Jarah Evslin} 
\affiliation{Institute of Modern Physics, CAS. NanChangLu 509, Lanzhou 730000, China}
\affiliation{University of the Chinese Academy of Sciences, YuQuanLu 19A, Beijing 100049, China}
\author{Qiang Fu}
\affiliation{Institute of Modern Physics, CAS. NanChangLu 509, Lanzhou 730000, China}
\affiliation{University of the Chinese Academy of Sciences, YuQuanLu 19A, Beijing 100049, China}
\author{Jian Tang~\footnote{Corresponding author: tangjian5@mail.sysu.edu.cn}}
\affiliation{School of Physics, Sun Yat-Sen University, Guangzhou 510275, China}

\begin{abstract}
Coherent elastic neutrino-nucleus scattering (CE$\nu$NS) can be used to determine the neutron part of nuclear form factors, unlocking intrinsic properties of nuclear structure. In contrast with other such methods, CE$\nu$NS is free from both strong interaction effects and Coulomb distortions. We propose precision measurements of CE$\nu$NS with an upcoming accelerator facility and determine the corresponding requirements for such a neutrino detector. We find that most significant backgrounds come from fast neutrons, induced by cosmogenic muons or from the pion decays at rest in the target station. With ton-scale liquid noble gas detectors, we will not only achieve percent level precision in the measurement of neutron radii but also clarify contributions of higher-order moments to nuclear form factors.

\end{abstract}

\maketitle

\section{Introduction}

The Standard Model (SM) of particle physics has been known to predict coherent neutrino scattering since the 1970s~\cite{Freedman:1977xn}. The measurement of such coherent neutrino-nucleus interactions is a fantastic achievement, robustly testing the SM. Several on-going experiments and proposals have been realized since the first theoretical prediction. The TEXONO experiment uses GW-level reactor neutrino sources and High-Purity Germanium (HPGe) detectors to conduct searches~\cite{Kerman:2016jqp}. The CONNIE proposal plans to adopt similar neutrino sources with neutrino detection by low-threshold CCD detectors~\cite{Aguilar-Arevalo:2016qen}. The RED-100 experiment combines the reactor neutrinos and a two-phase LXe neutrino detector to search for CE$\nu$NS~\cite{Akimov:2012aya}. The COHERENT experiment~\cite{Bolozdynya:2012xv,Collar:2014lya,Akimov:2015nza} intends to make use of stopped pions at the Spallation Neutron Source and perform detections with a list of optional detectors, including technologies with Argon, HPGe, CsI and the like. Very recently the CsI detector working group has announced the first observation of coherent neutrino-nucleus scatterings at the COHERENT experiment~\cite{Akimov:2017ade}. 

The most robust method to measure the neutron distribution so far is polarized electron scattering, as proposed in Ref.~\cite{Donnelly:1989qs}. Polarized electrons are scattered off of nuclei and the total elastic cross section is measured for each polarization. Electromagnetic interactions lead to polarization independent cross sections and so the difference arises from electroweak interactions. The PREX collaboration performed such an experiment~\cite{Abrahamyan:2012gp}. The statistics were, however, compromised leading to only a 2$\sigma$ observation of the neutron skin of $^{208}$Pb. In Ref.~\cite{Horowitz:2012tj}, the neutron radius of Pb was calculated using the data from PREX, with a precision of around 3\%. Recently, using the first data from COHERENT, the neutron distributions of Cs and I were estimated with a precision of around 20\%~\cite{Cadeddu:2017etk}. Exchange of a single $Z^0$ is not the only process which violates parity. One may also exchange a $Z^0$ and a $\gamma$. If the nucleus remains in its ground state between the exchange of these two bosons, the correction to the cross section results in a Coulomb distortion, which is quite large as calculated in the case of $^{208}$Pb in Ref.~\cite{Horowitz:1998vv}. However, $^{208}$Pb is the heavy nucleus with the largest energy separation between its ground state and the first excited state. More generally, two boson exchange in which the nucleus passes through an excited state may need to be considered. Experience with two boson exchange in other settings~\cite{Blunden:2003sp,Arrington:2011dn} suggests that large theoretical uncertainties may result. 

A measurement of the neutron density distribution in a nucleus removes a key uncertainty in atomic physics. In Ref.~\cite{Pollock:1992mv}, it was already shown that the use of atomic parity violation to test the Standard Model with subpercent precision requires a determination of the neutron distribution which is more precise than the spread in the models available at the time. Therefore, a measurement of the neutron distribution, via atomic physics, may lead to discoveries in fundamental physics. Further implications of a measurement of the neutron density require some model dependence. While one expects in general that a higher pressure of the neutron equation of state implies a larger neutron radius and so a thicker neutron skin, there is in fact a linear relation between neutron skin thickness and neutron pressure in the case of Skyrme-Hartree-Fock models~\cite{Brown:2000pd} at $0.1$ neutrons/fm${}^3$. More nontrivially, the same linear relation applies to relativistic Hartee models~\cite{Typel:2001lcw}. Thus a model independent measurement of the thickness of the neutron skin would provide a reasonably robust determination of the pressure of bulk neutron-rich matter, such as is expected in the crust of a neutron star. This pressure in turn determines the density and thickness of the crust itself~\cite{Horowitz:2000xj}.

{\tempbf{The observation of CE$\nu$NS opens new opportunities, which provide the motivation for this study.}} First, any observed deviation from the SM may indicate new physics. Many studies have focused on Non-Standard Interactions in coherent neutrino-nucleus scattering, such as Refs.~\cite{Lindner:2016wff, Dent:2016wcr, Coloma:2017egw, Shoemaker:2017lzs}. Second, neutrino-nucleus scattering cross sections and more generally collective neutrino behavior are critical inputs for the understanding of core collapse supernovas, which in turn is essential to understand supernova nucleosynthesis~\cite{Wilson:1974zz}. In Ref.~\cite{Centelles:2008vu}, authors claimed that measurements of neutron skins of a broad selection of nuclei can determine the nuclear symmetry energy in bulk nuclear matter. This was shown for a variety of mean field theory models with consistent results between models. Extrapolations to more neutron rich environments that may not be realized in the laboratory would allow a determination of energy functionals for neutron stars and various stages of core collapse supernovae. Continuous theoretical and experimental efforts can lead to a better understanding of coherent neutrino scattering with various nuclei, which will in turn be fed into supernova simulations. Third, direct detection of dark matter is rapidly approaching the intrinsic neutrino background floor~\cite{Billard:2013qya}. CE$\nu$NS events have to be well understood before we step on that floor. Detectors with different target nuclei require a clear understanding of the neutrino response. Actually, it is even possible to constrain the photon-portal Dark Matter directly with CE$\nu$NS~\cite{Ge:2017mcq}. Fourth, form factors from nuclear physics enter the game. A form factor is a transformation of the nucleon distribution. Traditionally, nucleon distributions of nuclei were determined by proton scattering. However, strong interaction uncertainties led to anomalies which were increasingly difficult to avoid \cite{Ray:1985yg} arising from initial state interactions and medium effects~\cite{Horowitz:1999fk}. Initial state interactions may be avoided using an electromagnetic probe. Pion photoproduction for example was recently used \cite{Tarbert:2013jze} to measure the neutron skin. However, the interaction of the pion and the nucleus again introduces strong interaction uncertainties arising from final state interactions and medium effects. Following the suggestion of Ref.~\cite{Tsoneva:2003gv}, electric dipole polarizability has been used as a probe of the neutron skin thickness \cite{Tamii:2011pv}. While the strong correlation between these two quantities is quite well motivated~\cite{Piekarewicz:2006ip}, an extrapolation of the neutron skin thickness from the polarizability has proven to be quite model dependent~\cite{Piekarewicz:2012pp}. In the case of neutrino scattering, on the other hand, $\gamma$ exchange is forbidden and two $Z$ exchange is suppressed by several orders of magnitude. While the first generation of neutrino scattering experiments may not yield the lowest uncertainties, they will be the most free from such theoretical uncertainties, in particular in the case of nuclei with relatively low lying excited states. The CE$\nu$NS cross section is dominated by the number of neutrons in the target nucleus. Precision measurements of CE$\nu$NS will help to extract robust information about nucleon distributions of nuclei. 

Unpolarized electron scattering already provides a precise measurement of electromagnetic nuclear form factors and, at low momentum transfer $Q$, the nuclear density distribution of protons. Neutron distributions in nuclei have already been measured, in many cases with quoted uncertainties which are less than those which may be attained with the setups proposed here. However, all measurements performed so far involved either electromagnetic or strong interactions with the target nucleus, leading to additional uncertainties which may be difficult to quantify. On the other hand, neutrinos only interact via electroweak interactions and so provide the cleanest probe of nuclear structure, in the sense that the uncertainties themselves will be well known. CE$\nu$NS is complimentary to existing measurement techniques. Combining two scattering techniques towards a measurement of the neutron density, one arrives at a model-independent determination of the thickness of the proton or neutron skin.

\tempbf{A CE$\nu$NS facility requires a dedicated neutrino source.  Worldwide, there are several high intensity accelerator facilities which could host pion decay-at-rest beam lines, including the Spallation Neutron Source in the US where the COHERENT experiment is running, the 590 MeV Ring Cyclotron at the Paul Scherrer Institute, the European Spallation Neutron Source (ESS) being built in Sweden, the 3 GeV proton beamline at the Japan Proton Accelerator Research Complex and the China Spallation Neutron Source (CSNS).  We will consider CSNS in our physics study of nuclear form factor measurements using CE$\nu$NS. } 

This article is organized as follows:  
in Section~\ref{sec:cenns}, we focus on the possibility of measuring neutrino-nucleus coherent cross sections using pion decay-at-rest neutrinos and explore potential detection techniques to conduct CE$\nu$NS in order to improve our understanding of nuclear structure in the low-energy range. The expected precision of form factor measurements is presented in Section~\ref{sec:sensitivity}. Finally, we will summarize our results and outlook in Section~\ref{sec:summary}.

\section{Formalism and detection of CE$\nu$NS}
\label{sec:cenns}

For a nucleus at rest with Z protons and N neutrons, the differential cross section for coherent neutrino-nucleus scattering is~\cite{Scholberg:2005qs}:
\begin{equation} \label{eqn:XS}
 \frac{d\sigma (E_\nu\,,E_r)}{d E_r} = \frac{G_F^2 [N- (1-4\sin^2\theta_w)Z]^2 F^2(Q^2) M^2}{4 \pi}\times \frac{1}{M} \left(1-\frac{E_r}{E_{max}}\right)
\end{equation}
where $G_F$ is the Fermi coupling constant, $\theta_w$ is the weak mixing angle, M is the mass of the nucleus and $E_r$ is the nuclear recoil energy. The maximum nuclear recoil energy, $E_{max}$, depends on the initial neutrino energy $E_\nu$ and the nucleus mass $M$:
\begin{equation}
 E_{max}=\frac{2E_\nu^2}{M + 2E_\nu}
\end{equation}
The coupling of the proton and $Z^0$ is proportional to $1-4{\rm{sin}}^2\left(\theta_w\right)$ which is coincidently very small because ${\rm{sin}}^2\left(\theta_W\right)\sim 0.23$ and the two terms nearly cancel each other. There is no such cancellation for the coupling of the neutron to the $Z^0$.  As a result, the neutron couples to the $Z^0$ an order of magnitude more strongly than the proton, and so the electroweak nuclear form factors are dominated by the neutron part of the nuclear form factor, which is the Fourier transform of the neutron density. Therefore, measurements of the electroweak nuclear form factors can be used to map the neutron density distribution of the nucleus. 

The form factor is a transformation of the density distribution:
\begin{equation}
 F(Q^2) = \frac{1}{Q_w}\int \left[\rho_n(r) - (1-4\sin^2\theta_w)\rho_p(r) \right] \frac{\sin(Qr)}{Qr}r^2 dr
 \label{eqn:FF}
\end{equation}
where $Q_w=N- (1-4\sin^2\theta_w)Z$ is the weak nuclear charge and $Q^2\simeq 2M E_{r}$ is the squared momentum transfer for CE$\nu$NS. 
Density functional theory, for example, predicts the form factors for different nuclear matter, surface and deformation properties~\cite{Patton:2012jr}.

Neutrino coherent scattering is rare. But the basic requirements for such searches are similar to those already used for the direct detection of dark matter. One always prefers a large fiducial mass, low threshold and extremely low radioactive backgrounds in an underground laboratory with good shielding from neutrons and cosmic muons. HPGe has the lowest threshold for neutrino detection at around 10 eV so far, and allows extremely low backgrounds in the target. However it is hard to scale up the total fiducial mass and deal with signal cross talk in the readout. A similar semi-conductor technology using CCDs has a higher threshold to identify neutrino scattering signals which limits the count rate. Novel technologies using gaseous proportional counters are under development. One such detector is the NEWS-G at LSM and SNOLAB~\cite{Arnaud:2017bjh}. On the other hand, liquid noble gas detectors are widely used in dark matter search experiments. The typical energy threshold of nuclear recoil for a liquid Xenon (LXe) or liquid Argon (LAr) detector approaches the sub-keV range. It seems plausible that liquid noble gas detectors may go beyond the current limit as several multi-ton detectors using LXe/LAr are in the queue.   Here we focus on LXe/LAr as the cross section of CE$\nu$NS is proportional to the number of neutrons in the target.  It is easy to generalize the current study to other types of detectors, though the sensitivity towards nuclear form factors highly depends on a range of neutrino beams, detector scenarios and background suppression techniques. 
\subsection{Neutrino fluxes}
\begin{figure}[!t]
 \includegraphics[width=0.45\textwidth]{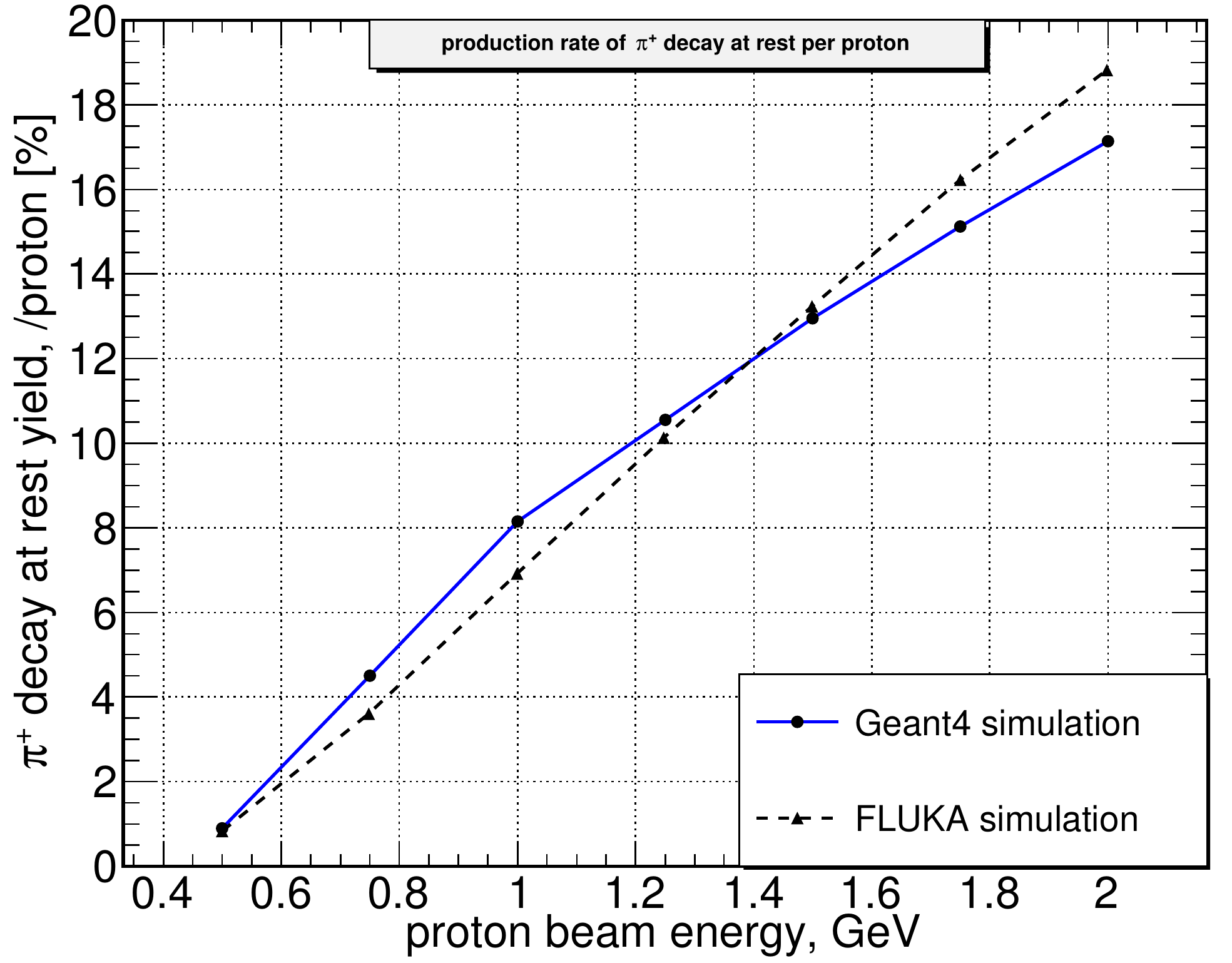}
 \caption{\label{fig:fluxes} Neutrino yields from pion decay-at-rest rates as a function of the proton energy cross checked with GEANT4 and FLUKA, considering a solid tungsten target.}
\end{figure}
The primary proton energy at the China Spallation Neutron Source (CSNS) is 1.6 GeV with a beam power of 100 kW during Phase I. The beam current in the rapid cycling synchrotron is 62.5 $\mu$A with a repetition rate of 25 Hz corresponding to a 40 ms time window. Each pulse has 2 bunches with a time structure of around 70 ns per bunch which implies a duty factor of $3\times10^{-6}$. One target station has been constructed mainly for neutron instruments. The number of protons per pulse is about $1.3\times10^{13}$. One of the beam lines is to be dedicated to muon physics studies. An Experimental Muon Source (EMuS) will be provided for the Muon Spectrometer (MuSR) and the development of beam preparations for the MOMENT experiment which we can use for accelerator neutrino oscillation experiments to detect CP violation and non-standard interactions in the leptonic sector~\cite{Cao:2014bea, Blennow:2015cmn, Bakhti:2016prn, Tang:2017qen}.  

EMuS is going to run in three different modes~\cite{Vassilopoulos:2017von}. In the neutrino mode, a wide energy spectrum of neutrinos will come from pion decays-in-flight, with the energy spread between 300 and 500 MeV. The surface muon mode produces pion decay-at-rest, yielding muon neutrinos with a momentum spread of $\pm 5$\% about 30 MeV. The third mode with muons decaying in flight offers a neutrino beam with a larger momentum spread: $\pm 10$\% in the energy range from 100 to 200 MeV. Phase II of CSNS, now under discussion, is an upgrade from 100 kW to 500 kW with one more target station. Similarly, the China Initiative Accelerator Driven System (CIADS), whose civil construction has just begun, will run a 500 MeV continuous wave proton beam with a 10 mA current in order to drive an experimental subcritical nuclear reactor~\cite{Li:2013bsa, Zhan:2017ayi}. It can offer a much higher current and make a precision measurement with CE$\nu$NS possible. However, the background will be considerably higher, since the beam is not pulsed. 

We consider the high-luminosity proton beam at CSNS. It will isotropically create muon neutrinos for CE$\nu$NS via $\pi$DAR. In Fig.~\ref{fig:fluxes}, we show the simulated neutrino yields or equivalently $\pi$DAR rates as a function of the proton beam energy and expressed as the ratio $\pi$DAR/proton. The simulation is cross checked with GEANT4 and FLUKA. Since we will always assume one-year lifetime in our simulations, the total number of neutrinos produced can be obtained by multiplying this number by the number of protons on target in one year. If the neutrinos produced at CIADS are used, the total flux will be about 9 times higher. However, the CIADS beam is not pulsed so that the steady state backgrounds will be considerably higher. It will require that a better veto system should be combined with the current proposed neutrino detector.

\subsection{Neutrino detection technology and background assumptions}

We choose sophisticated detection technology using liquid Argon (LAr) and liquid Xenon Time Projection Chambers (LXe TPC).
In a single-phase LAr detector, the particles that hit the target Argon atom with enough energy transfer will produce excited singlet and triplet states, followed by scintillation light at a wavelength of 128 nm when they decay to the ground state. The capture of this scintillation light constitutes the detection of the incoming particles. A typical nuclear recoil (NR) creates more singlet states than a typical electronic recoil (ER). The lifetimes of singlet and triplet states in LAr differ by three orders of magnitude. If we adopt light sensors to record waveforms of NR and ER, NR and ER will present very different shapes. Then a pulse shape discrimination (PSD) based on waveforms can be used to separate NRs and ERs.  Xenon also creates scintillation light when particles pass through the detector. Lifetimes for singlet and triplet states, however, are of the same order of magnitude and so PSD is less effective for xenon detectors. Fortunately, we can rescue the separation of NRs and ERs using a combination of the scintillation signal (S1) in the liquid phase and the ionized electron signal (S2) in the gas phase. High voltage has to be applied in order to drift the ionized electrons from the pure liquid Xenon to the gas. \tempbf{The drift time for the ionized electrons is of order $\mathcal{O}(100) \mu$s~\footnote{The long drift time of S2 signal in LXe TPC might be a potential problem to suppress the cosmic-induced backgrounds if we only use the timing structure in the beam. The passive shielding veto has to be considered.}. In a liquid nobel gas detector, extensive calibration processes have to be conducted in order to correctly connect the nuclear recoil energy to the light yield. Especially for a LXe TPC, we need the Noble Element Simulation Technique (NEST) to model the generation of scintillation photons and ionization electrons~\cite{Szydagis:2011tk, Lenardo:2014cva}.}

\begin{figure}[!t]
 \includegraphics[width=0.44\textwidth]{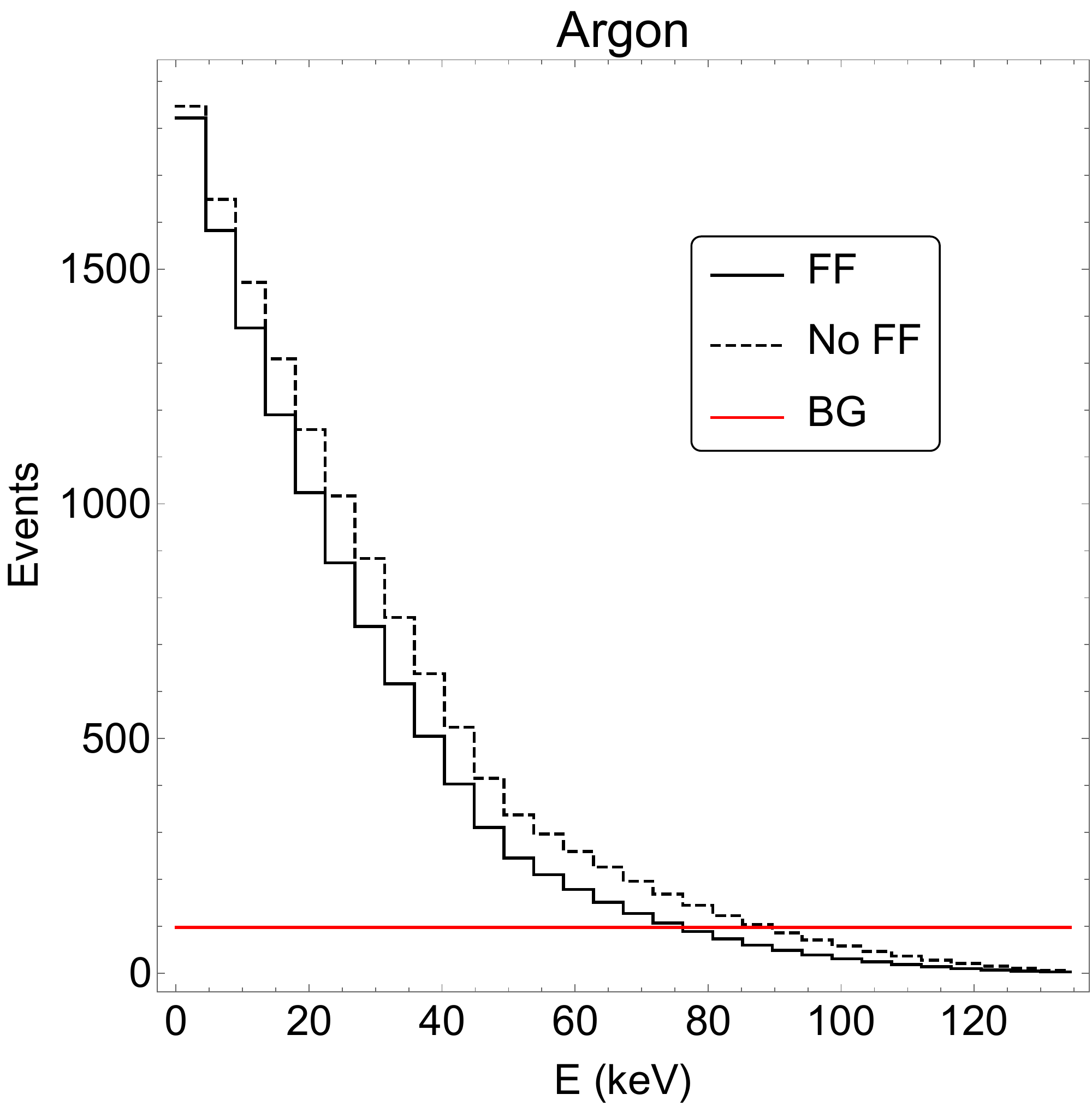}
 \includegraphics[width=0.45\textwidth]{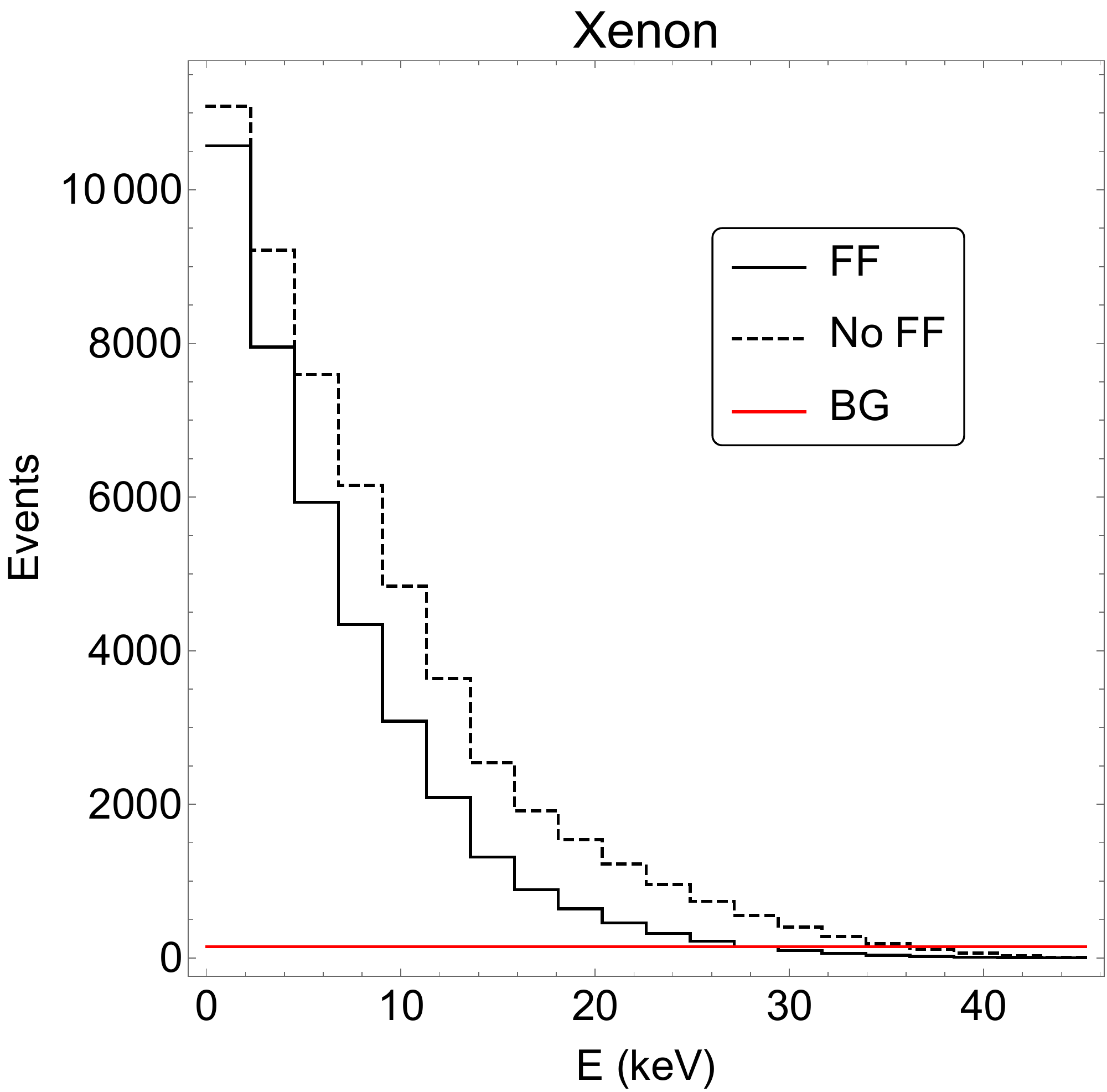}
 \caption{\label{fig:spectra} Spectra of CE$\nu$NS signals with/without form factors taken into account labelled with thick black line and dashed black line, respectively. The beam-correlated backgrounds are flat and labelled as steady-state backgrounds. Spectra for a liquid Argon and Xenon detector running for one year are shown in the left and right panel, separately.}
\end{figure}

Here we list the major backgrounds for a CE$\nu$NS detector and techniques needed to increase the signal to noise ratio. 
\begin{description}
 \item [Beam-correlated backgrounds] {
 Fast neutrons coming from the target station will provide nuclear recoils in the neutrino detector as we share the proton beam from a neutron spallation source. A proper shielding in the detector design and calibration is needed to suppress them efficiently. Other residues from the target station could be misidentified as neutrino signals since the neutrino detector is close to get enough statistics. We suppose they can be subtracted by background modelling and the timing structure of the pulsed proton beam.
 }
 \item [Beam-uncorrelated backgrounds] {The typical backgrounds in the direct detection of Dark Matter (DM) can be expected here, including those induced by cosmogenic muons and radioactive decays of $^{238}$U, $^{232}$Th and $^{40}$K in detector materials and the surrounding environment. Generally speaking, strict selections of low-background materials will be adopted at such a rare-event detector. In addition, fiducial volume cuts after position reconstruction will help to reduce the contamination by radioactive decays. LXe TPC allows a superb precision for position reconstructions. In a LAr detector, it seems that beta decays of $^{39}$Ar will dominate scintillation light by electromagnetic interactions. We, however, learned from experiments like DEAP-1 and DEAP-3600 using LAr where the PSD method provides a suppression factor of $\sim10^{-8}$ to identify nuclear recoils among the overwhelmingly dominant electromagnetic recoils~\cite{Amaudruz:2016qqa, Amaudruz:2017ekt}. We believe that it is not a problem for PSD to distinguish the CE$\nu$NS signal from $^{39}$Ar beta decays. For a LXe TPC detector, it is essential to combine direct scintillation signals in the liquid phase with escaped electron signals in the gas phase in order to pin down contamination from electromagnetic recoils, which was recently realized by XENON1T at the level of $(1.93\pm0.25)\times10^{-4}$ events/(kg$\times$day$\times$keV$_{ee}$)~\cite{Aprile:2017iyp}. \tempbf{For the cosmogenic muon-induced fast neutron backgrounds, one requires passive shielding using lead or water, because the timing structure in the beam might not be good enough to suppress these backgrounds and the detector is likely to be only a few meters below the surface.  With no shielding, the overwhelming muon-induced neutrons would result in event misidentifications. Similar conceptual designs to suppress muon-induced neutrons have been considered using LAr and LXe TPC detectors~\cite{Brice:2013fwa,Akimov:2017hee}.} }
\end{description}

\begin{figure}[!t]
 \includegraphics[width=0.44\textwidth]{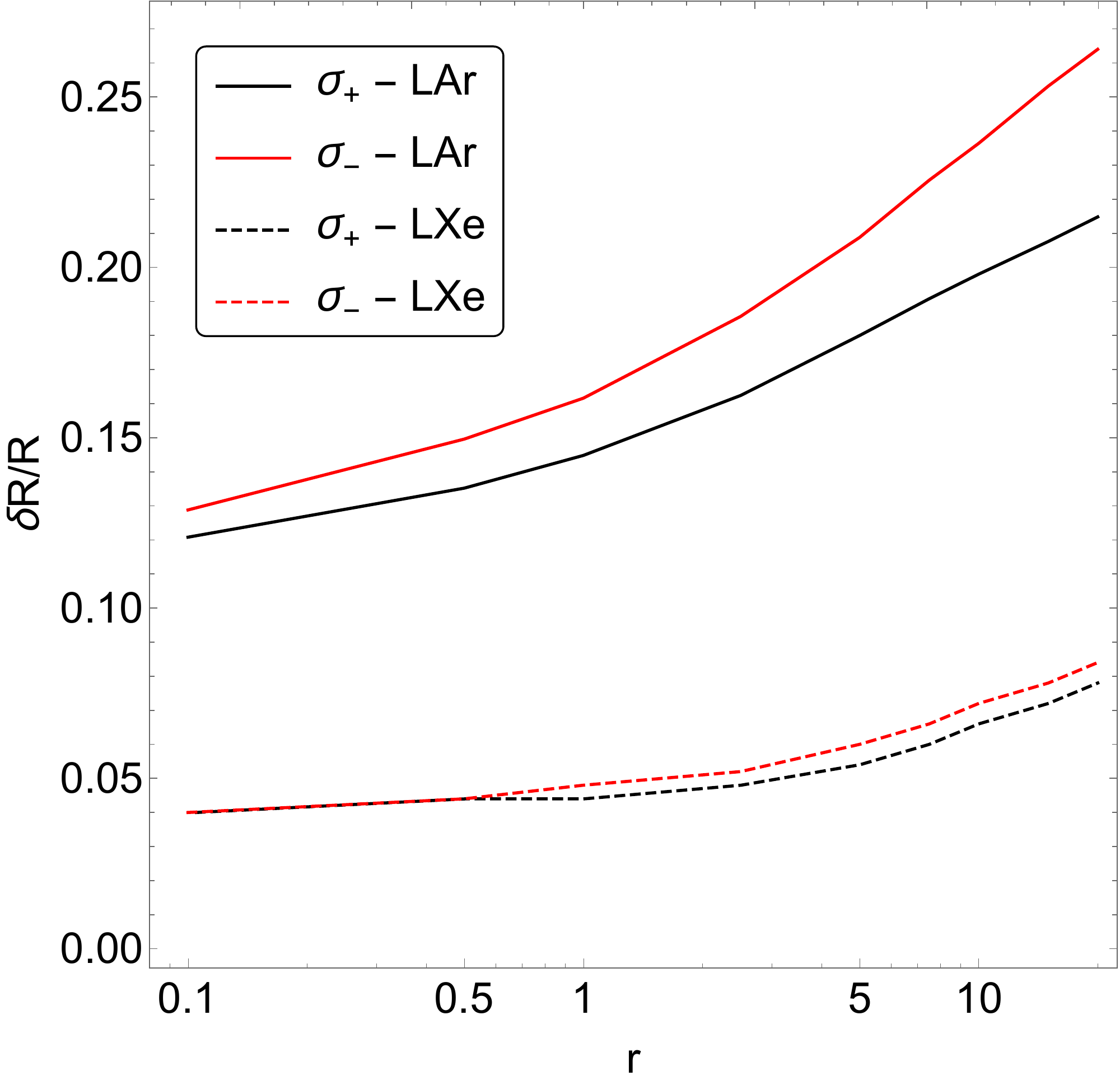}
 \caption{\label{fig:background} Sensitivity for 1 ton LAr and LXe detectors, assuming different background fluxes (1 year lifetime). {\tempbf{$\sigma_{+(-)}$ indicates the 1-$\sigma$ upper (lower) bound, as defined in Eqn.~(\ref{eqSigma})}}.  For each datapoint, the background rate is multiplied by a factor $r$ ($r=1$ corresponds to 3000 background events)}
\end{figure}

We consider the following baseline configuration in our simulations: a one-ton fiducial mass for LAr and LXe TPCs; an energy threshold of $10$ keV for the LAr detector and $3$ keV for the LXe TPC. The exact spectrum and total number of events of the background depends on many factors, such as the shielding, the position of the detector, etc.  We estimated the total number of background events for a LXe detector using Ref. \cite{Akimov:2015nza}, since the overburden considered is similar to what can be realistically assumed to be the one at CSNS; after taking into account the different detector mass and background reduction factor, we found that the expected number of steady-state background events for a 1 ton LXe detector at the CSNS facility is around 3000 events/year; we assumed a similar rate also for the LAr detector. We made a strong assumption regarding the background spectrum, considering a flat distribution. A detailed study of backgrounds will be the subject of a complete simulation with GEANT4 in the near future.

In Fig.~\ref{fig:spectra}, we show event spectra and backgrounds in LAr and LXe detectors, respectively. Thanks to the significant prompt neutrino fluxes from pion decays-at-rest, it is relatively easy for us to identify clear signals despite the beam-correlated backgrounds and steady state backgrounds. In Fig.~\ref{fig:background} we show the effect on the expected sensitivity if we change the total number of background events, rescaling it by a factor $r$ ranging from $0.1$ to $20$.

\section{Sensitivity to nuclear form factors with CE$\nu$NS}
\label{sec:sensitivity}

\begin{figure}[!t]
 \includegraphics[width=0.44\textwidth]{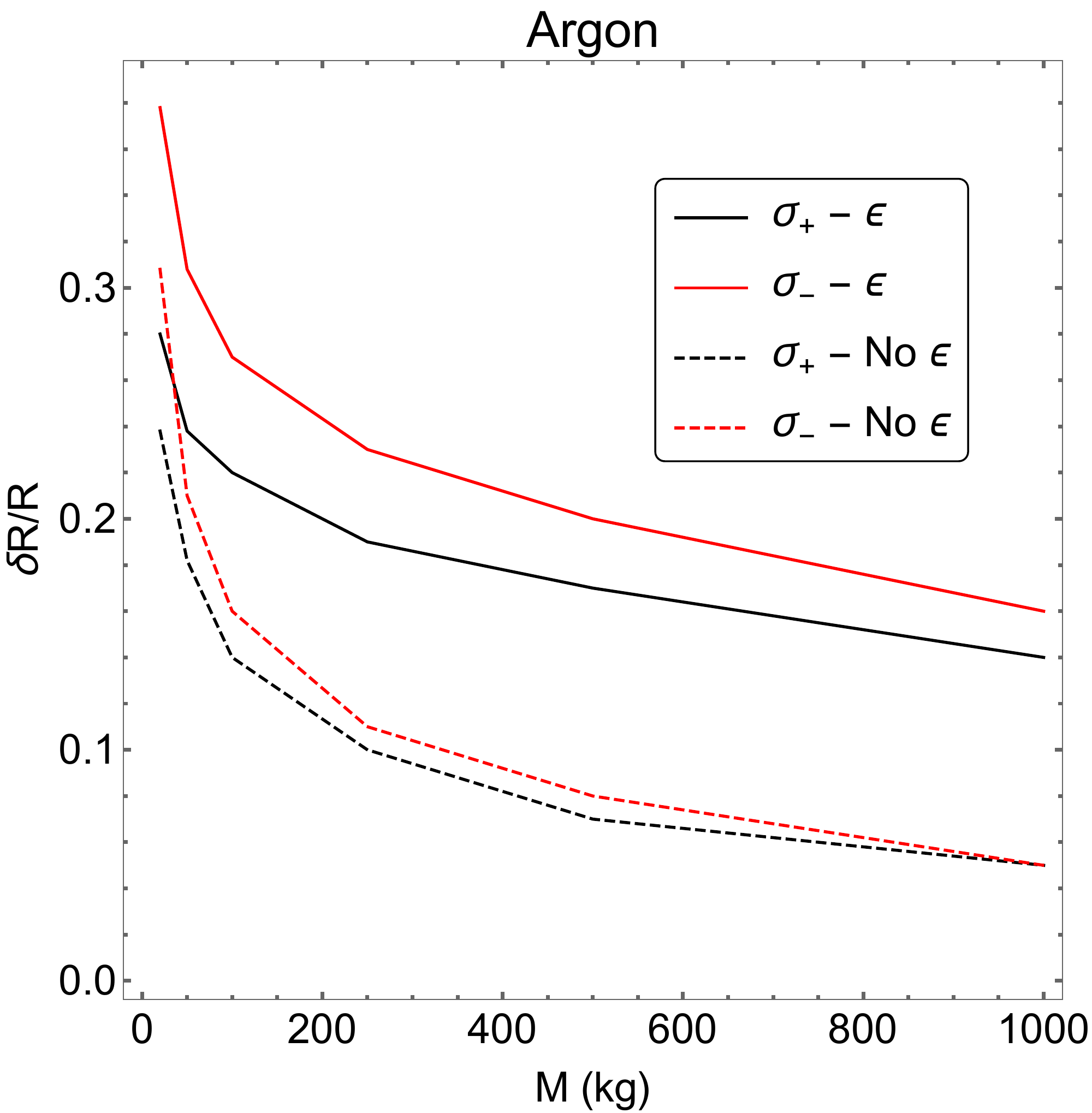}
 \includegraphics[width=0.45\textwidth]{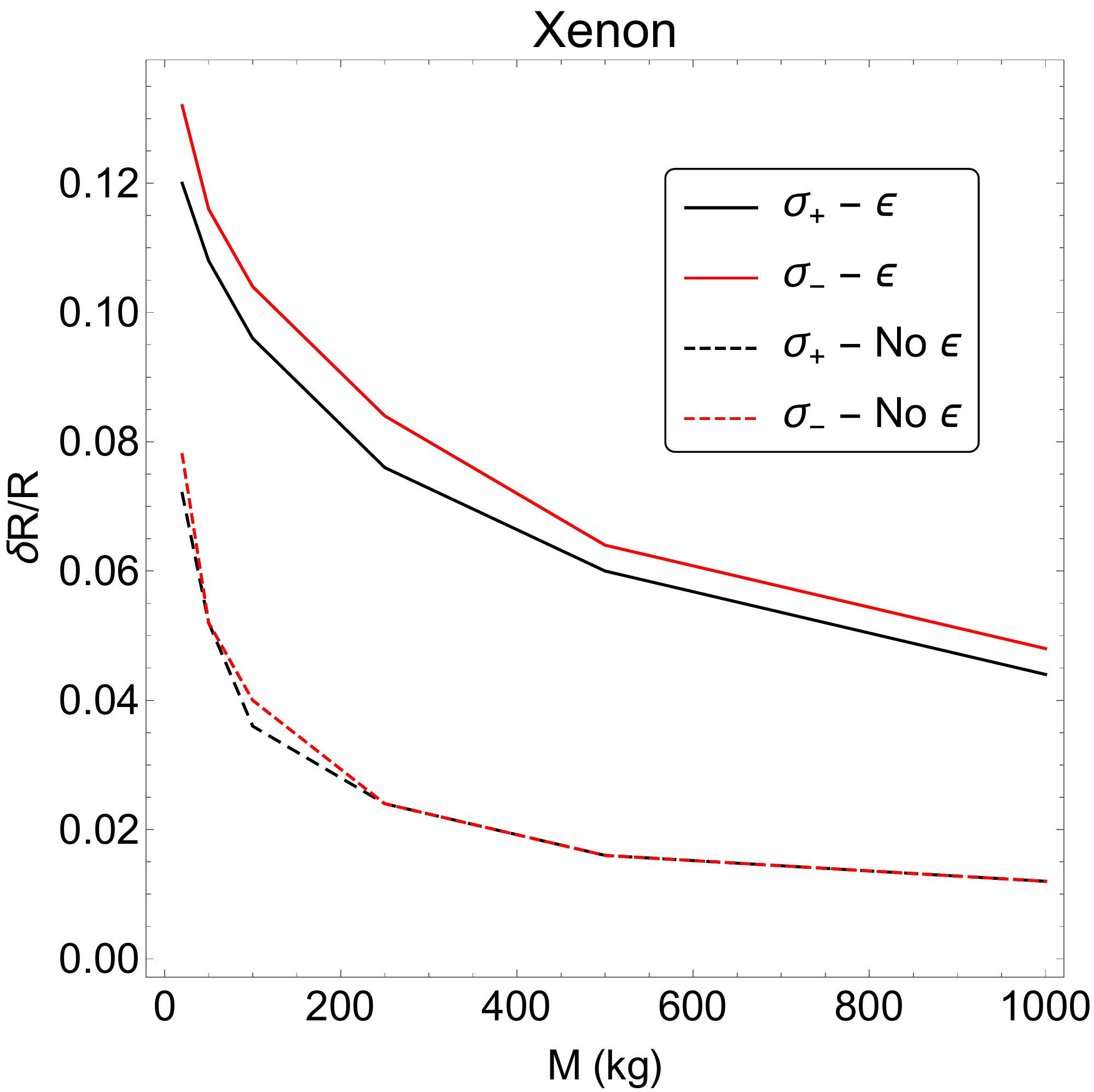}
 \caption{\label{fig:precision} Precision measurements of neutron parts of nuclear form factors at $1\sigma$ confidence level in terms of massive Liquid Argon and Xenon detectors, respectively, considering a 1 year lifetime. {\tempbf{ $\sigma_{+(-)}$ indicates the 1-$\sigma$ upper (lower) bound, as defined in Eqn.~(\ref{eqSigma})}}. Here $\epsilon$ represents the pull parameter related to the uncertainty on the quenching factor. The dashed curves represent cases minimized over $\epsilon$. The solid curves: $\epsilon$ is kept constant and equal to 0.}
\end{figure}

Compared with reactor neutrinos, neutrinos from $\pi$DAR have higher energies and so a higher tolerance to the threshold, thus simplifying neutrino detection technology.  Timing structure from the pulsed proton beam can be used to suppress the accidental backgrounds significantly. The drawbacks include the neutron fluxes from the spallation process. Usually we have to properly design the shielding structure or optimize the distance between the target station and the detector location. 
We will determine the sensitivity to nuclear form factors. First of all, we use the Helm model to describe the neutron distribution inside the nucleus~\cite{Helm:1956zz}. In the Helm model, the form factors depend on the diffraction radius $R_0$ and the surface thickness $s$. The impact of $s$ is negligible in the considered energy range. In all of the following calculations, $s$ is considered fixed and equal to 1 fm. The diffraction radius is related to the neutron distribution radius $R_n$ using the relation~\cite{Engel:1991wq}
\begin{equation}
 R_n^2=R_0^2+5s^2
\end{equation}
where $\langle R^2\rangle=3/5R_n^2$, $\langle R^2\rangle$ being the second moment of the neutron distribution.

The task is then to conduct precision measurements to determine the neutron distribution radius $R_n$. We quantify the precision that can be achieved for a given experimental configuration estimating the 1-$\sigma$ region, defined using the equation
\begin{equation}\label{eqSigma}
\Delta\chi^2(R_{bf}\pm\delta R_{\pm})=\chi^2(R_{bf}\pm\delta R_{\pm})-\chi^2(R_{bf})=1
\end{equation}

\begin{figure}[!t]
 \includegraphics[width=0.45\textwidth]{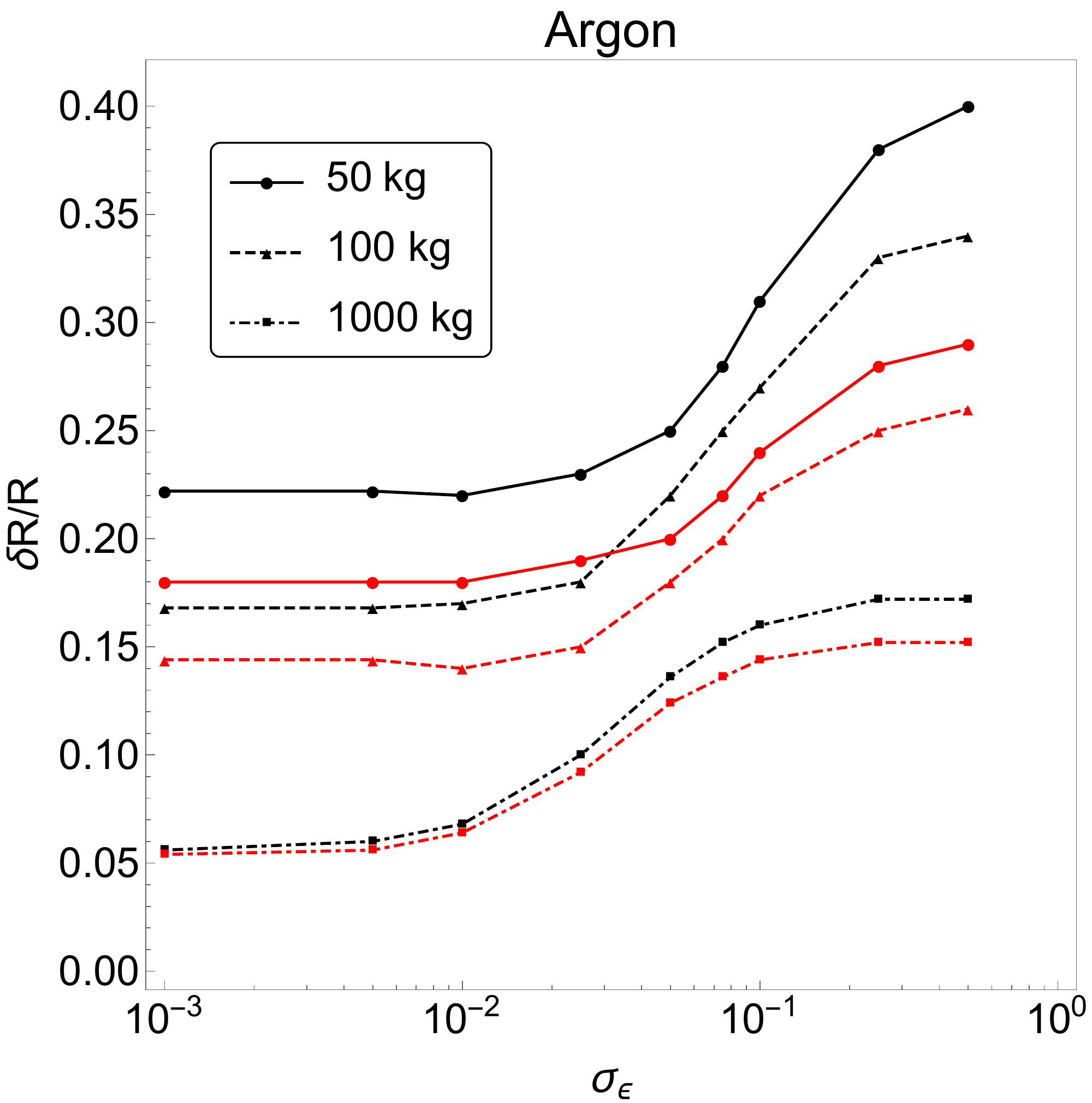}
  \includegraphics[width=0.45\textwidth]{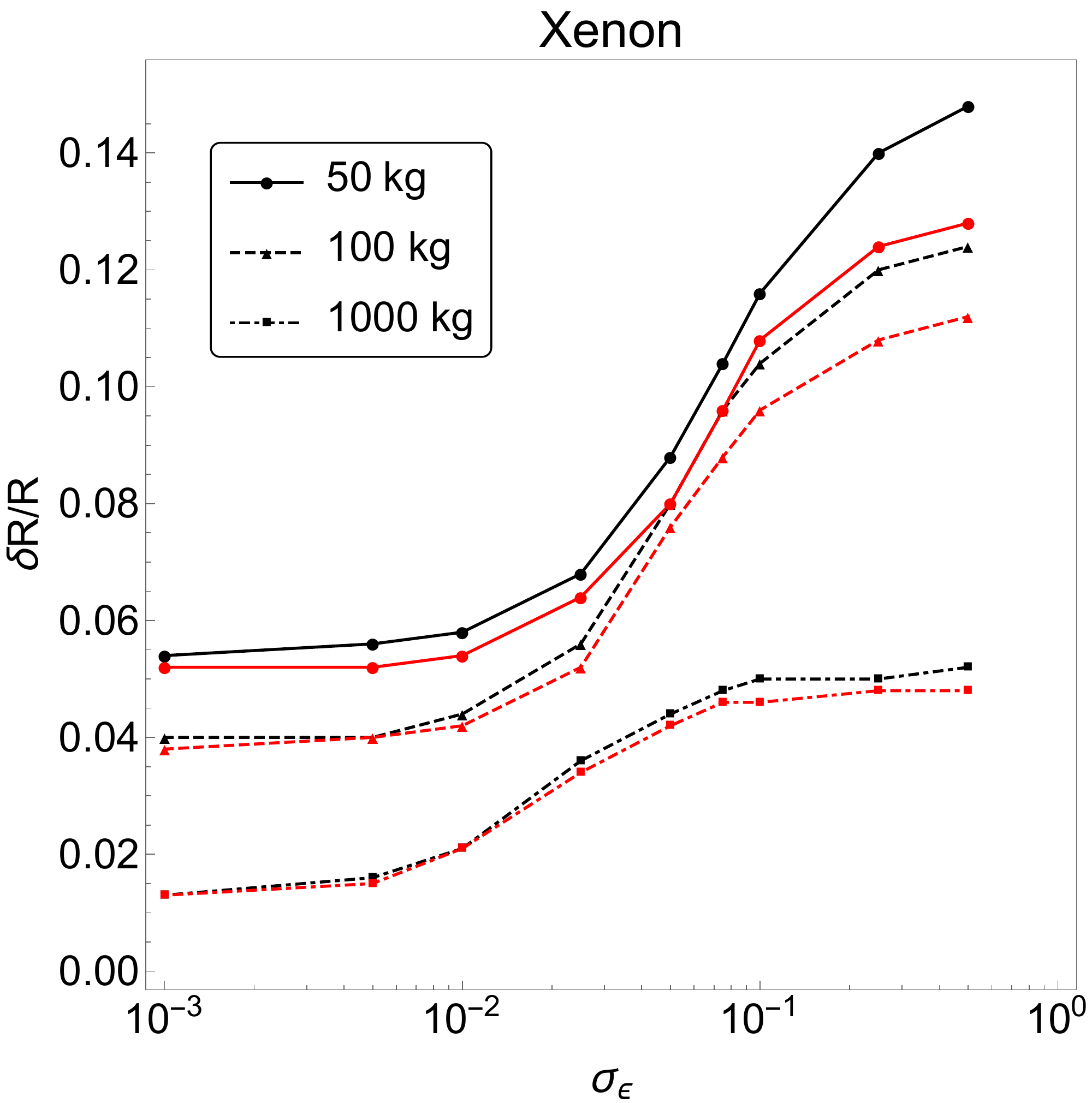}
 \caption{\label{fig:varieEpsilon} 1-$\sigma$ upper (black curves) and lower (red curves) bounds for Argon (left panel) and Xenon (right panel) detectors, as a function of the uncertainty on the quenching factor, for 3 different detector masses with one year data taken.}
\end{figure}

where $R_{bf}$ is the best-fit radius and $\delta R_{\pm}$ define the 1-$\sigma$ upper and lower bounds. {\tempbf{In these calculations we employed the Asimov data set method~\cite{Cowan:2010js}.  In particular the 1-$\sigma$ region was estimated using the asymptotic distribution ({\it i.e.} the theoretical expected spectrum).  As a consequence, $R_{bf}=R^{det}_n$, where $R^{det}_n$ is the value assumed for $R_n$ for Argon and Xenon (4.1 fm and 6.1 fm, respectively)}}. All of the pull parameters considered are minimized. In particular, a pull parameter $\alpha$ related to the total flux normalization is always used, with $\sigma_{\alpha}=0.1$. When noted below,  we introduce a pull parameter $\epsilon$ to quantify the uncertainty on the quenching factor, assuming that the observed energy $E_{obs}$ is given by
\begin{equation}
E_{obs}=E_{real}(1+\epsilon)
\end{equation}
where $E_{real}$ is the real recoil energy. If not otherwise specified, $\sigma_\epsilon$ is assumed to be 0.1. This reference value was chosen because it is roughly similar to the precision that is currently achieved by the COHERENT experiment~\cite{Akimov:2017ade}. However, it is reasonable to assume that it will be possible to improve the precision on the quenching factor in future experiments (later, we will also study the precision that can be achieved as a function of $\sigma_\epsilon$). {\tempbf{We present our results in Fig.~\ref{fig:precision}, reporting the dimensionless fraction $\delta R/R$.  Here one observes that, when the uncertainty on the quenching factor is considered, the precision that can be achieved is significantly worse:  with a 1 ton detector, it goes from 5\% to 15\% for an Argon detector and from 1.2\% to 4.6\% for a Xenon detector.}} Needless to say, a precise determination of the quenching factor is a prerequisite for a determination of the neutron part of the nuclear form factors.  

\begin{figure}[!t]
 \includegraphics[width=0.45\textwidth]{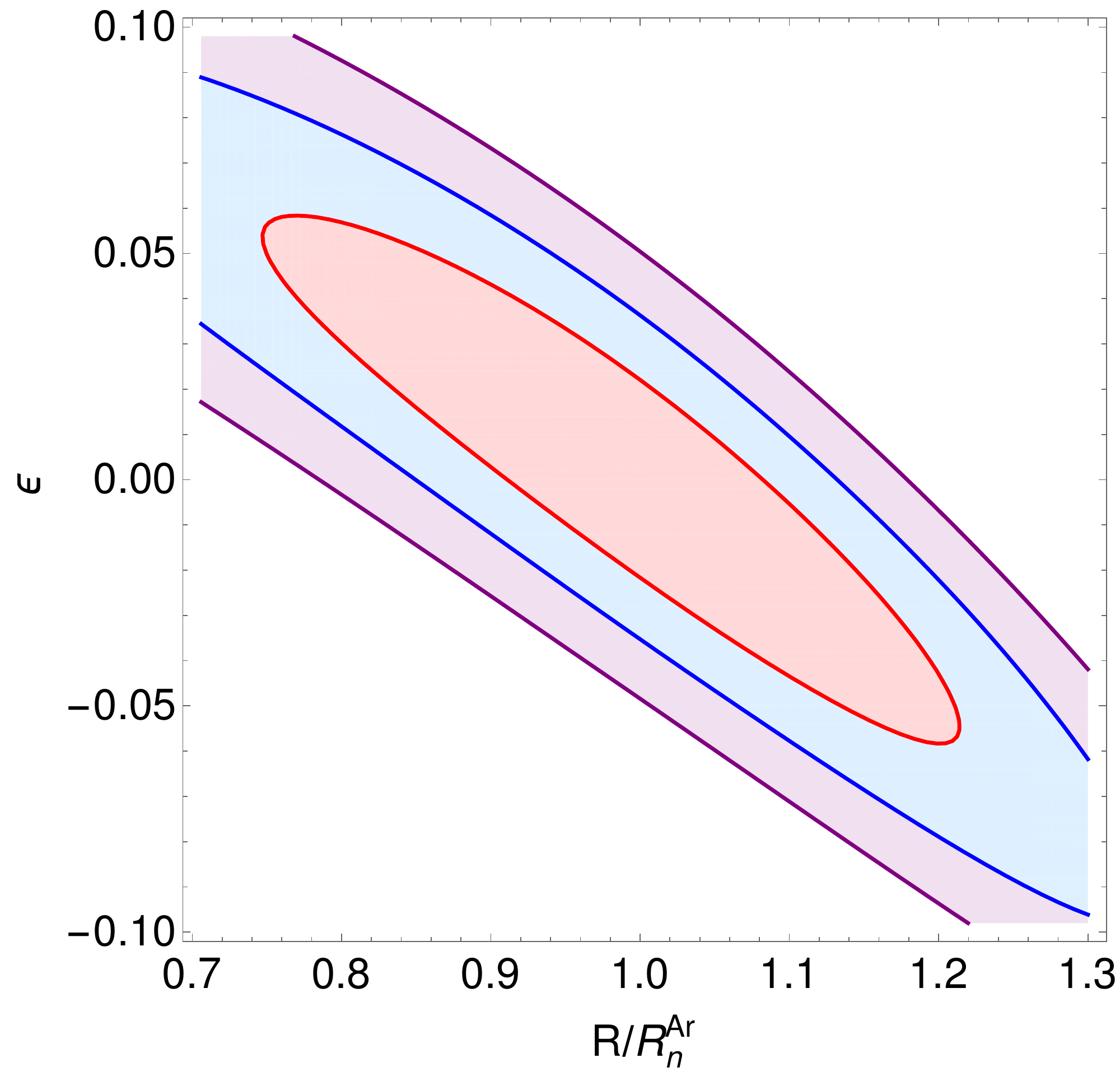}
 \caption{\label{fig:2Dplot} 1-$\sigma$, 2-$\sigma$ and 3-$\sigma$ regions (red, blue and purple curves, respectively), for a 1 ton LAr detector, considering one-year data taken.}
\end{figure}

The reason that LXe detectors can achieve a higher precision is that, since the cross section is proportional to $N^2$, using argon the expected number of events will be considerably lower. If $\Delta\chi^2$ is parabolic, $\delta R_+=\delta R_-$; however we can observe that this is not always the case, especially when the statistics are poor. In Fig.~\ref{fig:varieEpsilon} we present the precision that can be achieved as a function of the uncertainty on the quenching factor.  Either increasing this uncertainty or else decreasing the statistics (for example, decreasing the detector fiducial mass), will increase the asymmetry of $\Delta\chi^2$.  

The strong degeneracy between $R_n$ and $\epsilon$ can be appreciated in Fig.~\ref{fig:2Dplot}, where the 1-, 2- and 3-$\sigma$'s regions are shown in the $R_n-\epsilon$ plane (however, since in this case we are considering a two-degree-of-freedom chi-squared distribution, these regions are defined by $\Delta\chi^2=2.3,\ 6.18,\ 11.8$).

\begin{figure}[!t]
 \includegraphics[width=0.3\textwidth]{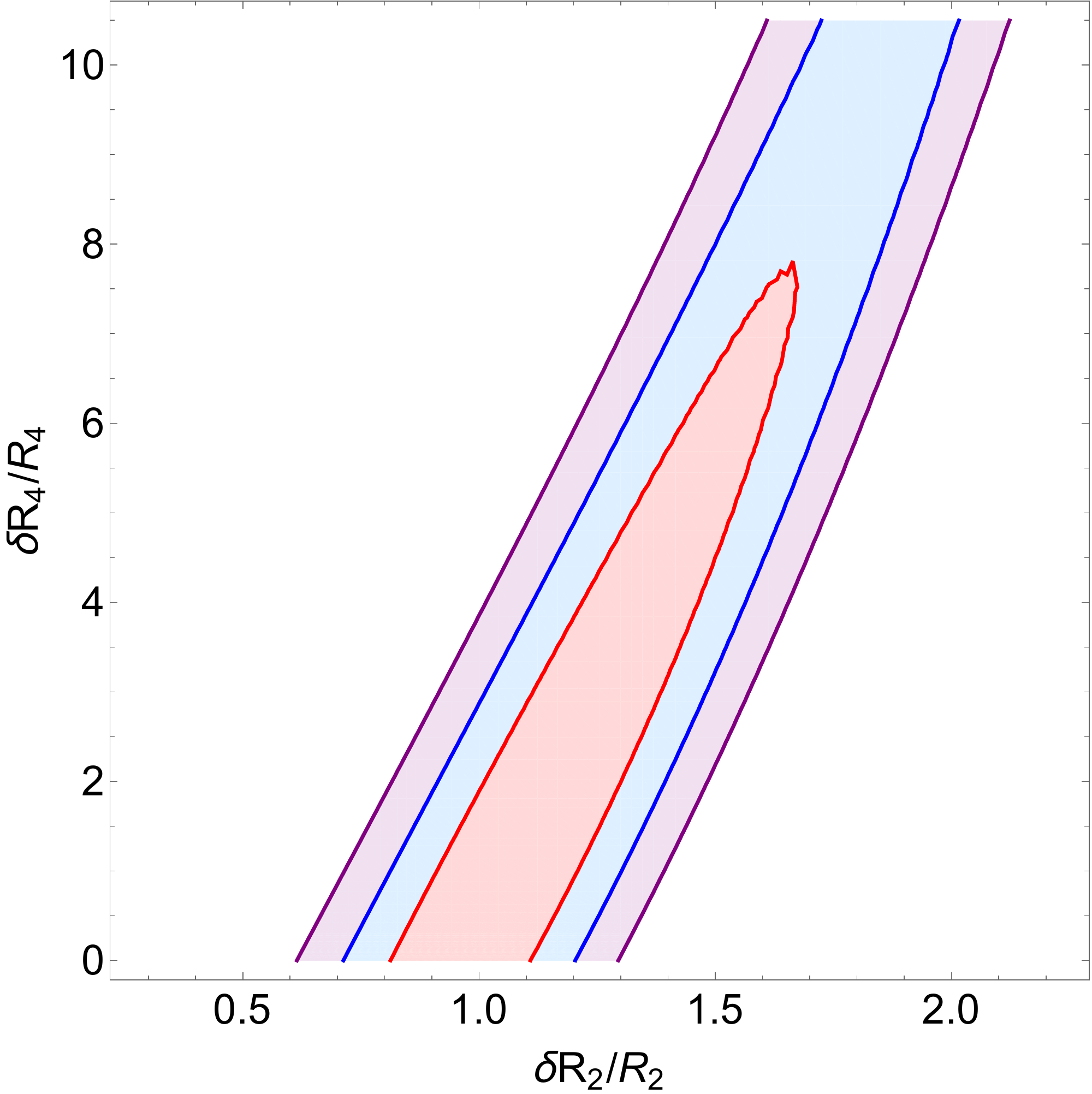}
 \includegraphics[width=0.3\textwidth]{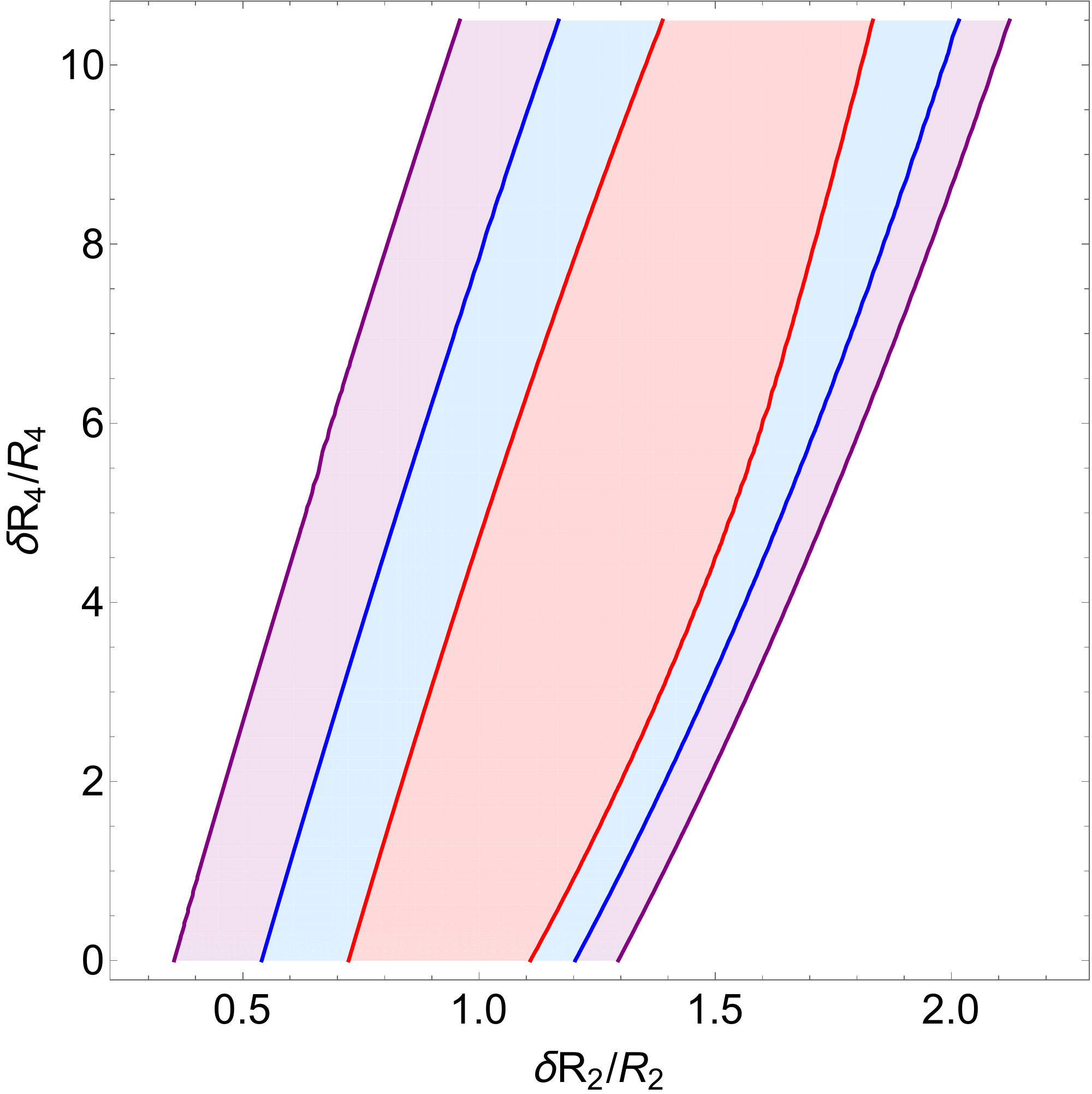}
 \includegraphics[width=0.3\textwidth]{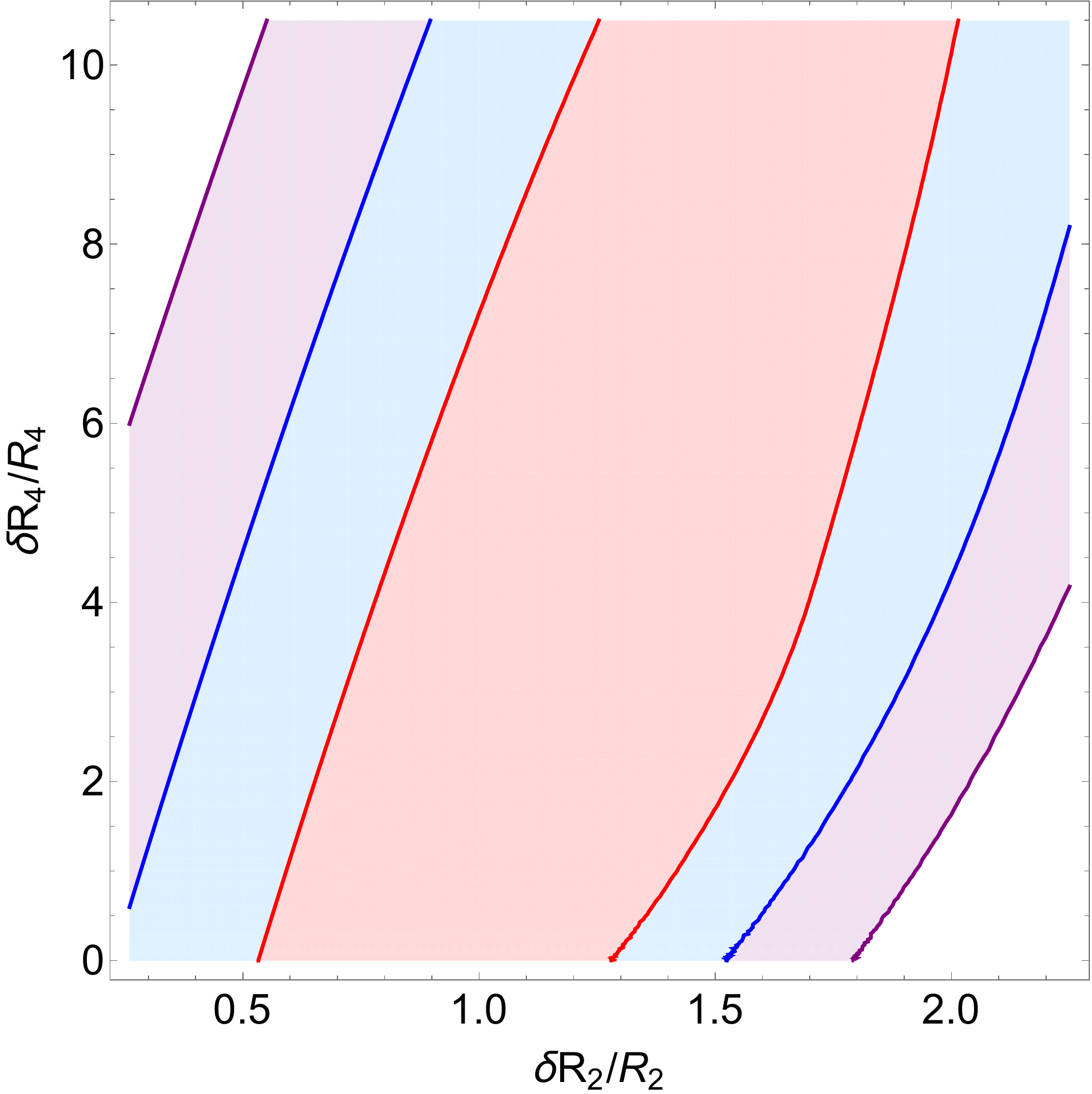}
 \caption{\label{fig:moments} Confidence regions of effective moments contributing to neutron parts of nuclear form factors at $1\sigma$(red), $2\sigma$(blue) and $3\sigma$(purple) confidence levels with a one-ton Liquid Argon detector and a 1 year lifetime. In the left panel, we marginalize over the flux normalization uncertainty only. The middle panel includes the second theoretical uncertainty due to truncated higher-order terms. The right panel includes the uncertainty from the quenching factor, assuming $\sigma_\epsilon=0.1$, {\it i.e.} roughly the same precision as is currently achieved by the COHERENT experiment.}
\end{figure}

As we reach a high-precision measurement, it is desirable to turn to a model-independent parametrization of the form factor.  A Taylor expansion in $Q^2$, as given in Eqn.~(\ref{eqn:FF}), allows for easy comparison with observables of interest. Each term $Q^{2n}$ will be multiplied by a coefficient proportional to the 2n-th moment of the neutron distribution, $\langle R^{2n}\rangle$.  We consider a one-ton LAr detector and expand the term $F(Q^2)^2$, given in Eqn.~(\ref{eqn:XS}), up to $\langle R^6\rangle$. In Fig.~\ref{fig:moments}, we observe the precision that we can achieve in the measurement of the neutron distribution moments, reported as the 1-, 2- and 3-$\sigma$'s regions in the $\langle R^2\rangle-\langle R^4\rangle$ plane. Here $\langle R^6 \rangle$, if not otherwise specified, is treated as a pull parameter and hence minimized. Information on the effective moments can be extracted and directly compared with theoretical predictions. 

\section{Summary}
\label{sec:summary}
{\tempbf{Several high-luminosity proton beams are either under construction or have been proposed.}} Neutrinos can be prepared via traditional fixed-target collisions which are widely used in accelerator neutrino oscillation experiments. In the meantime, the detector evolution in the direct detection of dark matter has tended towards the low-mass region with low-threshold detectors. {\tempbf{As a result, it is now for the first time possible to use CE$\nu$NS as a tool to study nuclear structure.  We consider a setup to conduct precision measurements of CE$\nu$NS at a $\pi$DAR facility, a relatively small experiment to bridge particle and nuclear physics. With ton-scale LAr and LXe detectors, one can achieve percent level precision on a measurement of the neutron radii and approach the determination of effective moments in the form factor expansion.}} We have shown that a low uncertainty on the quenching factor must be reached in order to achieve a good precision at this kind of measurement. CE$\nu$NS may not yet be the most precise tool to measure the neutron parts of nuclear form factors, {\tempbf{but it is quite robust and also sufficiently precise to distinguish and to test nuclear physics models.}}

\section{Acknowledgement}
This work is supported in part by the National Natural Science Foundation of China under Grant No. 11505301.   EC is supported by NSFC Grants No. 11605247 and 11375201, and by the Chinese Academy of Sciences Presidents International Fellowship Initiative Grant No. 2015PM063.  JE is supported by NSFC grant 11375201 and the CAS Key Research Program of Frontier Sciences grant QYZDY-SSW-SLH006.  JE and EC thank the Recruitment Program of High-end Foreign Experts for support.  JT acknowledges Kai Zuber for pointing out the typo in slides at the workshop of Jinping Neutrino Experiment in 2017, where Neutron Activation Analysis was accidentally labelled as the Neutrino Activation Analysis. This triggers the current study. We benefitted a lot from the workshop of MOMENT and EMuS held in SYSU on August 29 and 30, 2017. EC would like to thank Carlo Giunti for the useful discussions and comments.

\appendix
\section{Low-Energy Threshold}

\begin{figure}[h]
 \includegraphics[width=0.45\textwidth]{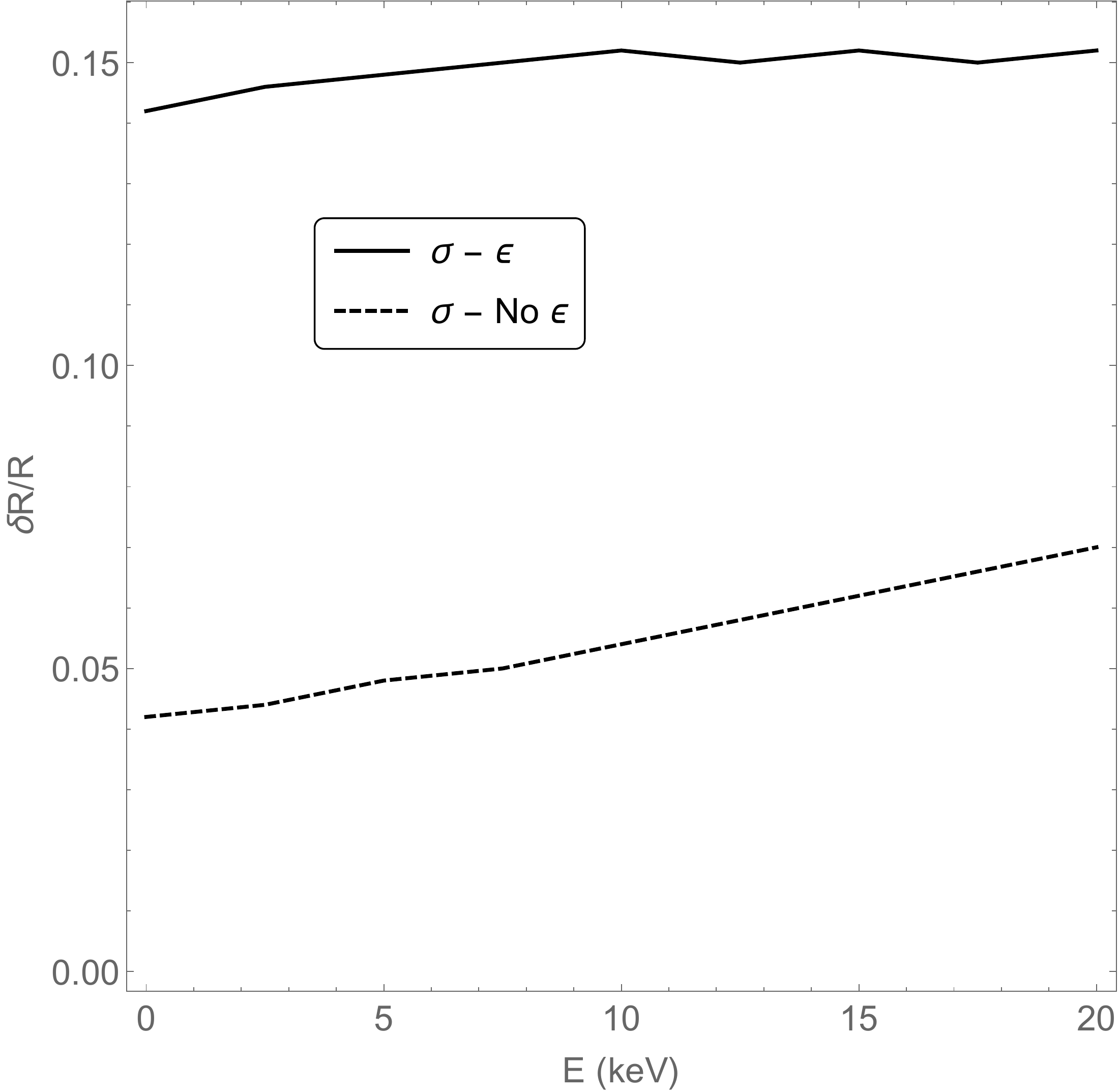}
 \caption{\label{fig:threshold} Sensitivity as a function of the low-energy threshold, considering a 1-ton LAr detector and one-year data taken.}
\end{figure}

The impact of a detector's energy threshold on the precision of the form factor measurement is not very significant with the current experimental setups. While most of the events are in the low-energy region (and hence a high threshold could considerably decrease the total number of CE$\nu$NS events observed), the deviation from full coherence is larger at higher energies, as can be seen from Fig.~\ref{fig:spectra}. This means that most of the information on the form factor can only be obtained from an observation of the high-energy tail of the reconstructed CE$\nu$NS spectrum. In Fig.~\ref{fig:threshold}, the expected precision is shown as a function of the detector energy threshold.

\bibliography{ref.bib}

\end{document}